\documentclass[aps,reprint,showpacs,showkeys,groupedaddress]{revtex4-1}

\usepackage{amsmath}
\usepackage{bm}
\usepackage{caption}
\usepackage{color}
\usepackage{float}
\usepackage{graphicx}
\usepackage{mathtools}
\usepackage{multirow}
\usepackage{subcaption}

\DeclareMathOperator\asinh{arcsinh}

\begin{document}


\title{Polarization-dependent excitons and plasmon activity in nodal-line semimetal ZrSiS}


\author{Juan J. Mel\'endez}
\affiliation{Department of Physics, University of Extremadura}
\affiliation{Institute for Advanced Scientific Computing of Extremadura \\ Avda. de Elvas, s/n, 06006 Badajoz, Spain}

\author{Andr\'es Cantarero}
\affiliation{Molecular Science Institute, Universitat de Val\`encia, PO Box 22085, 46071 Valencia, Spain}


\begin{abstract}
The optical properties of bulk ZrSiS nodal-line semimetal are theoretically studied within a many-body formalism. The $G_0W_0$ bands are similar to those calculated within the density functional theory, except near the $\Gamma$-point; in particular, no significant differences are found around the Fermi energy. On the other hand, the solution of the Bethe-Salpeter equation reveals a significant excitonic activity, mostly as dark excitons which appear in a wide energy range. Bright excitons, on the contrary, are less numerous, but their location and intensity depend greatly on the polarization of the incident electric field, as the absorption coefficient itself does. The binding energy of these excitons correlate well with their spatial distribution functions. In any case, a good agreement with available experimental data for absorption/reflection is achieved. Finally, the possible activation of plasma oscillations is investigated. Plasmons may be formed at low energies, but they are damped and decay producing electron-hole pairs, more importantly for $\bm q$ along the $\Gamma$-$M$ path.
\end{abstract}


\maketitle

\section{Introduction}\label{intro}
Since the discovery of their nontrivial topological behavior, the family ZrSiX, with X = S, Se or Te, have received considerable attention as archetypes of nodal-line semimetals (NLSMs) \cite{Neupane-16}. These are particular forms of Dirac or Weyl semimetals in which band cross along continuous lines within the first Brillouin zone \cite{Burkov-11}, the crossing being protected by non-trivial topological invariants. This fact gives NLSMs a set of exceptional optical and transport properties, which have been reviewed elsewhere \cite{Gao-19}. The most prominent member of the ZrSiX family, as also the most widely studied from the experimental and theoretical approaches, is probably ZrSiS just because of its outstanding physical properties. Strictly, nodal lines at ZrSiS appear at about 0.5-0.7 eV below the Fermi energy, where Dirac points protected by the non-symmorphic symmetries exist along some directions of the first Brillouin zone (BZ). A second set of quasi nodal-lines, slightly above the Fermi energy, are actually gapped because of the spin-orbit coupling (SOC), although the gap is small, of the order of tens of meV \cite{Schoop-16}. From these set of quasi nodes, the linear regime of the ZrSiS bands extends down to 2~eV below the Fermi energy, which is larger than in any other compound known to date. 

The band structure and non-trivial topology of NLSMs is responsible for the outstanding physical properties of these systems. In this context, ZrSiS exhibits a large and strongly anisotropic magnetorresistance, which is consistent with the existence of two Fermi valleys at the valence band at the Fermi level and depends drastically on temperature \cite{Singha-17,Sankar-17}. In particular, for $\bm H \parallel \bm c$ up to 9~T, the magnitorresistance along [100] reaches 8500\% at 3 K and 1.4$\cdot10^5$ at 2 K, but decreases to about 14\% at 300 K. Besides, this system features a high mobility of the (hole-dominated) charge carriers \cite{Sankar-17}, reaching 3$\cdot 10^4$ cm$^2$V$^{-1}$s$^{-1}$ at 3 K, with potential interest for thermoelectric applications under magnetic fields \cite{Matusiak-17}. Finally, ZrSiS exhibits an intense Zeeman effect at low magnetic fields \cite{Hu-17}. This fact arises from the combination of the aforementioned high charge carriers mobility with a nearly zero effective mass, and results in an exceptionally large Land\'e gyromagnetic factor ($\approx$ 38). The other members of the ZrSiX family seem to exhibit interesting physical properties as well, although they have been much more scarcely studied so far\cite{Hu-16,Ebad-19,Gatti-21}. 

Optical properties of NLSMs have attracted the attention of the scientific community as well\cite{Weber-18}. In this respect, it was early discovered that ZrSiS features a U-shaped optical conductivity, with a plateau in the infrared ending at a sharp peak at around 1.3~eV; alternatively, the absorption coefficient takes on very low values below about 1~eV, and it increases sharply at higher energies\cite{Ebad-19,Schilling-17}. This fact, which is common to other NLSMs, is surprising, since the optical conductivity of a electron-hole symmetric Dirac (or Weyl) semimetal in $d$ dimensions is expected to vary as $\sigma_1(\omega) \propto \omega^{(d-2)/z}$, where $z$ is an exponent describing the curvature of the bands near the gap\cite{Bacsi-13}. This behavior is well known in graphene, for which $\sigma_1 \approx \text{constant}$, or in 3D systems such as TaAs or Cd$_3$As$_2$, for which $\sigma_1(\omega) \propto \omega$ \cite{Xu-16,Neubauer-16}. In NLSMs, the existence of Dirac bands causes 2D electrons to exist in the three-dimensional bulk so that, somehow, these 3D materials behave as graphene at some energy range. This is a very simplistic explanation to the ``flat'' region of the $\sigma_1(\omega)$ curve of ZrSiS, though. In fact, the observed trend for the optical conductivity has been theoretically explained in terms of the non-linearity of the low-energy band structure by a tight-binding model\cite{Habe-18}. 

In any case, the optical properties of ZrSiS seem to be affected by correlation effects in the electron-hole interactions. Actually, Huh et al. have shown that the screening of the Coulomb interaction is only partial when a nodal line exists parallel to the Fermi surface\cite{Huh-16}. In this case, the nodal line does not lie exactly on the Fermi energy (and therefore the DOS does not exactly vanish at $\varepsilon_F$), but still one may expect a long-ranged electrostatic interaction between electrons and holes. In particular, it has been theorized that the ground state of a NLSM with congruent electron and hole Fermi surfaces becomes unstable in presence of electron-hole interactions (see \cite{Rudenko-18,Scherer-18} and references therein), in a manner that recalls the Cooper instability in a superconductor. The electron-hole pairing (or, more generally, the correlation effects between charge carriers) provides a physical explanation to the experimental evidence of mass enhancement of charge carriers in ZrSiS\cite{Pezzini-18}, and the question arises as to whether the optical properties of this system may be influenced by electron-hole interactions as well. 

Surprisingly, the theoretical studies about optical properties of ZrSiS are scarce, to our knowledge. Rudenko et al.\cite{Rudenko-18} use a tight-binding approach to suggest that a (pseudo)gap should appear in ZrSiS at low enough temperatures due to correlations between charge carriers; in particular, this could provide a physical explanation to the noticeable mass enhancement of charge carriers found by Pezzini et al.\cite{Pezzini-18}. On the other hand, Zhou et al. \cite{Zhou-20} used an independent-particle (IP) framework based on the density functional theory (DFT) and showed that the optical properties of ZrSiS, at this level of theory, are not affected much by uniaxial compression (up to stress of 10 GPa), except by the narrowing of the flat region with increasing strain. Interestingly, the IP calculations suggest that plasmon modes should exist at two energy ranges, namely at $\approx$ 5 eV (lossy plasmons) and $\approx$ 20 eV (stable plasmons), these ranges also being stable under stress. 

All the previous facts suggest that further insights onto the optical properties of ZrSiS are required, particularly those related to the effect of charge carrier correlations. In this paper, we have computed the dielectric function of ZrSiS by solving the Bethe-Salpeter equation (BSE) for electron-hole pairs in ZrSiS, with the focus on the calculations of optical coefficients and on the possible existence of plasmon modes. The obtained results compare fairly well with experimental data available, and provide a theoretical description for the behavior of charge carriers of ZrSiS from a many-body formalism. In addition, we confirm the existence of plasmons at low energies, which get dumped by decay onto electron-hole pairs. Finally, we confirm that hyperbolic plasmons, which have been predicted from the IP approach, may exist as well.

\section{Computational details}
\begin{figure}[!htb]
	\centering
	\begin{subfigure}[b]{0.45\textwidth}
		\centering
		\includegraphics[width=0.55\columnwidth]{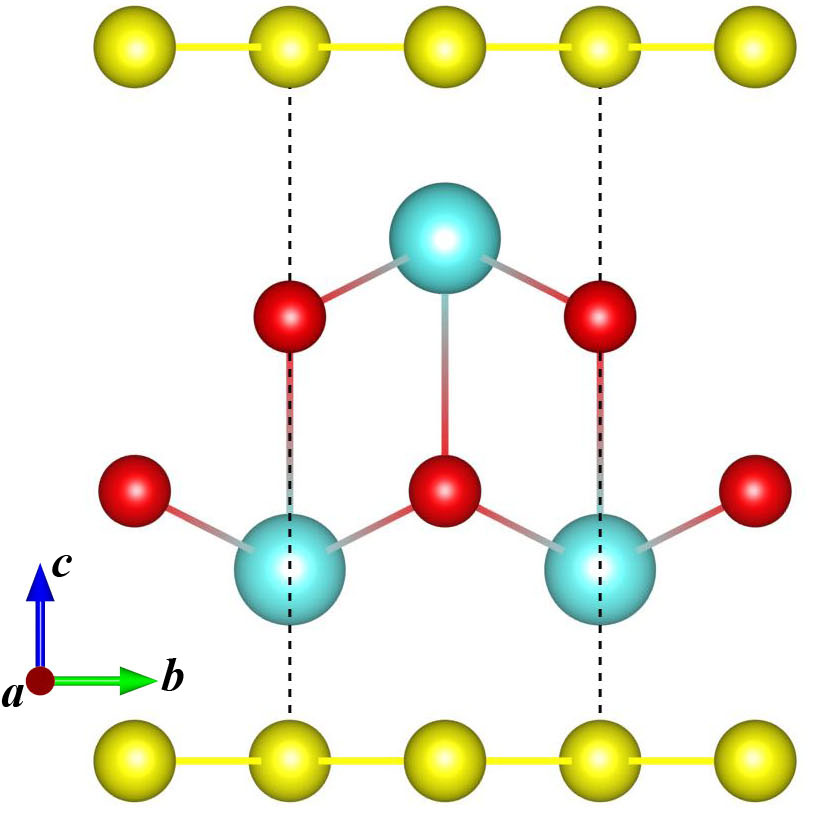}
		\label{subfig:unit_cell_a}]
	\end{subfigure}
	
	\vspace{\baselineskip}

	\begin{subfigure}[b]{0.45\textwidth}
		\centering
		\includegraphics[width=0.55\columnwidth]{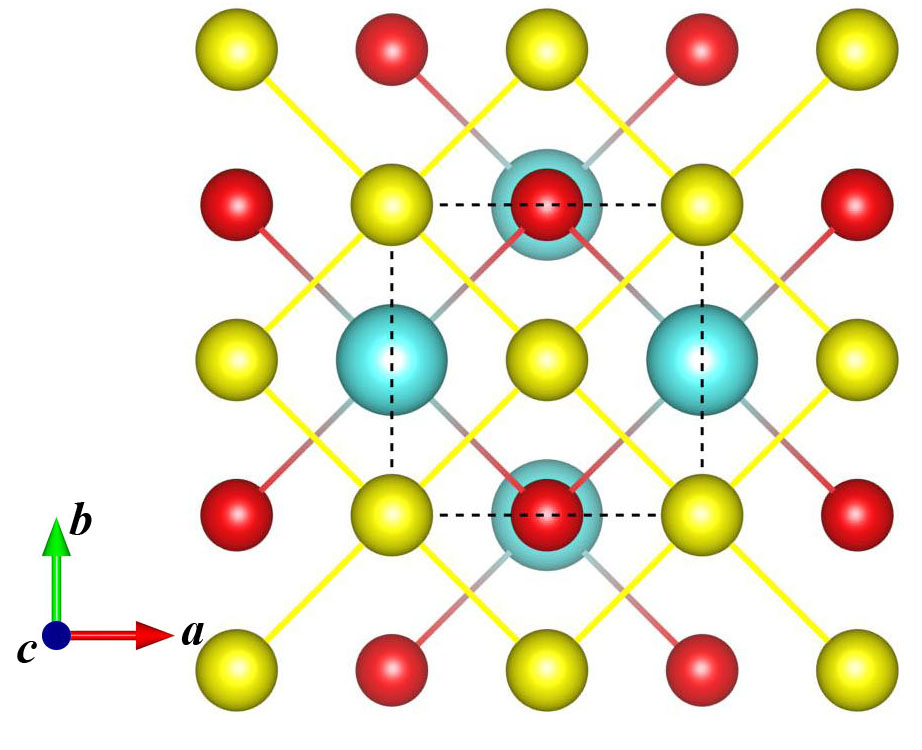}
		\label{subfig:unit_cell_c}
	\end{subfigure}
	\caption{Views of the ZrSiS unit cell along the $\bm a$ (\textit a) and $\bm c$ (\textit b) axes. The polarizations of the electric field mentioned below are referred to the cell axes shown in this figure.}
	\label{fig:unit_cell}
\end{figure}

The electronic structure of ZrSiS was first calculated from standard DFT within the generalized-gradient approximation for the exchange-correlation functional \cite{Perdew-96}; SOC corrections were taken into account explicitly by using full relativistic pseudopotentials for all the chemical species. The kinetic energy cutoff was set to 900 eV, and the BZ was sampled with a 21$\times$21$\times$7 $\bm k-$mesh centered at $\Gamma$. The structure of ZrSiS is shown in Figures~\ref{fig:unit_cell} as projections of the unit cell along the $\bm a$ and $\bm c$ axes; these axes define the electric field polarizations within this paper.

Starting from the experimental lattice parameters $a =$~3.5450(5)~\AA, $c =$ 8.0578(16)~\AA\ \cite{Schoop-16}, the structure was relaxed until the Hellmann-Feynman forces decreased below 3$\cdot$10$^{-6}$ eV/\AA; the relaxed lattice parameters where $a =$ 3.52~\AA\ and $c =$ 8.01~\AA, which compare fairly well with the starting ones. The DFT calculations were performed with the Quantum Espresso suite \cite{Giannozzi-09}. The DFT bands along high-symmetry directions of the BZ were interpolated from the maximally-localized Wannier functions, as implemented in the Wannier90 code \cite{Marzari-97,Souza-01}. 

The many-body corrections were accounted for within the $G_0W_0$ approximation \cite{Hedin-65} by using 961 $\bm G$-vectors for exchange. The dielectric function at the $G_0W_0$ level was computed within the plasmon-pole approximation \cite{Larson-13} including 200 bands and 201 $\bm G-$vectors for a converged value, and 400 bands were used in the sum defining the Green's function. The $G_0W_0$ energies were then used to build the BSE kernel, whose diagonalization provides the excitonic energies $E_\lambda$ and eigenstates $|\lambda \rangle$. From these, one may compute the component of the exciton in terms of the DFT electron-hole pair $A_{cv\bm k}^{\lambda}$ as
\begin{equation}
	A_{cv\bm k}^\lambda = \langle c v \bm k|\lambda\rangle,
	\label{eq:exciton_components}
\end{equation}
where $|c v \bm k\rangle$ denotes the DFT electron-hole state, namely the $v$-th state of the valence band and the $c$-th state of the conduction one at point $\bm k$. Given these variables, the macroscopic dielectric function is then computed as
\begin{widetext}
\begin{equation}
	\varepsilon_M(\bm q, \omega) = 1 - \frac {8\pi}{q^2\Omega} \sum_{vc\bm k} \sum_{v'c'\bm k'} \langle v, \bm k - \bm q| e^{-i\bm q\cdot \bm r} |c,\bm k\rangle \langle c', \bm {k'} | e^{i\bm q\cdot \bm r} |v', \bm {k'} - \bm q \rangle \sum_{\lambda} \frac {A_{cv\bm k}^{\lambda}A_{c'v'\bm {k'}}^{\lambda *}}{\omega - E_{\lambda}},
	\label{eq:epsilon}
\end{equation} 
\end{widetext}
where $\Omega$ is the unit cell volume, $v$ and $c$ are the number of valence and conduction bands which define the energy range of interest, $\bm k$ and $\bm k'$ are the DFT $\bm k$-vectors and the sum in $\lambda$ spans all the exciton states. In this work, the BSE equation was solved for four valence and conduction bands, which suffice to describe the low-energy region (i.e., $\varepsilon \lesssim$ 6 eV) of the optical spectrum. Describing the high-energy region ($\varepsilon \lesssim$ 20 eV), on the other hand, requires to consider transitions between deep valence and high conduction bands, which is well beyond our computational resources. Optical properties are calculated from Eq. (\ref{eq:epsilon}) in the limit $\bm q \rightarrow \bm 0$, with $\bm q$ parallel to the incident electric field. The contribution of the intraband transitions to the static screening matrix, needed to build and to diagonalize the BSE kernel, was modeled by a correction to the dielectric function of the form $\varepsilon^{\text{intra}}(\omega) = \varepsilon_1^{\text{intra}}(\omega) + i\varepsilon_2^{\text{intra}}(\omega)$, with

\begin{equation}
\begin{gathered}
	\varepsilon_1^{\text{intra}}(\omega) = \lim_{\delta \rightarrow 0^+}\left(1 - \frac {\omega_D^2}{\omega^2+\delta^2}\right) \\
	\varepsilon_2^{\text{intra}}(\omega) = \lim_{\delta \rightarrow 0^+}\frac {\delta \omega_D^2}{\omega^3 + \omega\delta^2},
\end{gathered}
	\label{eq:drude_correction}
\end{equation}
where $\omega_D$ is the Drude plasma frequency. In this case, we chose a polarization-dependent $\omega_D$, namely $\hbar\omega_D=1.088$~eV and $\hbar\omega_D=3.15$~eV for the $Ox$ and $Oz$ polarizations, respectively, which were reported to describe accurately the optical behavior at the IP level \cite{Zhou-20}. The imaginary part of the Drude plasmon was set to 0.005~eV. 

From the solutions of the G$_0$W$_0$ and the Bethe-Salpeter equations two additional useful functions may be computed. First is the electron energy loss function $L(\bm q, \omega)$, defined as
\begin{equation}
	L(\bm q, \omega) = -\text{Im }\left(\frac 1{\varepsilon_M(\bm q,\omega)}\right),
	\label{eq:def_L}
\end{equation}
where $\varepsilon_M(\bm q, \omega)$ is the macroscopic dielectric function \eqref{eq:epsilon}. The second relevant function is the finite-momentum joint density of states (fm-jDOS), $j(\bm q, \omega)$, defined as the number of states per unit energy separated by momentum $\hbar \bm q$ and energy $\hbar\omega$. This function may be calculated as
\begin{equation}
	j(\bm q, \omega) = \frac 1{\pi}\,\text{Im }\chi(\bm q, \omega),
	\label{eq:def_j}
\end{equation}
where $\chi(\bm q, \omega)$ is the complex susceptibility of the solid, which may computed from the $G_0W_0$ results. High $L(\bm q, \omega)$ values are indicative that energy is being lost at $\omega$, as a consequence of any inelastic mechanism. On the other hand, the fm-jDOS is directly related to the decay rate for a plasmon in the state $(\bm q, \omega)$ to electron-hole pairs satisfying energy and momentum conservation \cite{Mahan-00} and, therefore, high fm-jDOS values denote high generation rates for electron-hole pairs. The coupling of a plasmon to the charge carriers of a solid is a many-body problem whose lowest order corresponds to the coupling mediated by the electronic polarizability $\chi$ through a given coupling matrix element $g$. In particular, the decay rate for the plasmon to an electron-hole pair is given by \cite{Mahan-00}
\begin{widetext}
\begin{equation}
	\Gamma(\bm q, \omega) = \frac {2\pi}{\hbar} \lim_{\eta \rightarrow 0^+} \sum_{n,n',\bm k} |g_{n,n',\bm k}(\bm q,\omega)|^2 \frac 1{\pi} \text{ Im} \left[\frac {f_{n,\bm k}-f_{n',\bm k + \bm q}}{\hbar\omega - (\varepsilon_{n',\bm k + \bm q} - \varepsilon_{n,\bm k}) - i\eta}\right], 
	\label{eq:sup_4}
\end{equation}
\end{widetext}
where $\varepsilon_{n,\bm k}$ and $\varepsilon_{n',\bm k+\bm q}$ are the energies of the hole and electron states, respectively, $f_{n,\bm k}$ and $f_{n',\bm k + \bm q}$ are the respective equilibrium occupations and the sum spans to states whose momenta and energies differ by $\bm q$ and $\hbar\omega$, respectively. 

The matrix elements $g_{n,n',\bm k}(\bm q,\omega)$ for the coupling between plasmons and electrons were derived elsewhere, in the limit of weak coupling, from the Landau damping of the electron gas \cite{Pines-62} as $g_{n,n',\bm k}(\bm q, \omega) = 2\pi e^2\hbar\omega/q^2$, thus independent of $n, n'$ and $\bm k$. Therefore, Equation~(\ref{eq:sup_4}) reduces to 
\begin{equation}
	\Gamma(\bm q, \omega) = \frac {4\pi^2 e^2}{\hbar} \frac {\hbar\omega}{q^2} j(\bm q, \omega),
\end{equation}
where
\begin{widetext}
\begin{equation}
	j(\bm q, \omega) = \frac 1{\pi} \lim_{\eta \rightarrow 0^+} \text{ Im} \left[ \sum_{n,n',\bm k} \frac {f_{n,\bm k}-f_{n',\bm k + \bm q}}{\hbar\omega - (\varepsilon_{n',\bm k + \bm q} - \varepsilon_{n,\bm k}) - i\eta}\right] \equiv \frac 1{\pi}\text{ Im } \chi(\bm q, \omega),
	\label{eq:sup_5}
\end{equation}
\end{widetext}
as written in Equation~\eqref{eq:def_j} above. The $G_0W_0$ and BSE calculations were performed with the Yambo code \cite{Marini-09,Sangalli-19}; the interpolation of the $G_0W_0$ bands was carried out with Wannier90, as for DFT.



\section{Results and discussion}
\subsection{Band structure}
\begin{figure}[!htb]
	\centering
	\includegraphics[width=0.9\columnwidth]{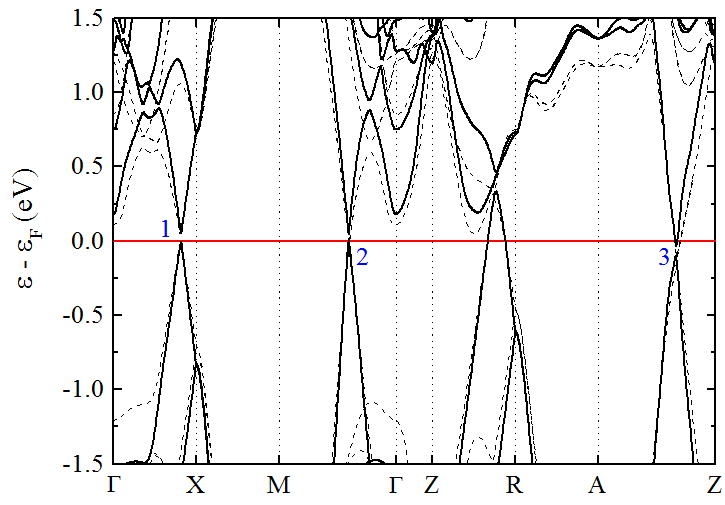}
	\caption{DFT (dashed lines) and $G_0W_0$ (solid ones) band structures near the gap of ZrSiS.}
	\label{fig:gw_bands}
\end{figure}

Figure~\ref{fig:gw_bands} displays the band structure of ZrSiS calculated at the DFT and $G_0W_0$ levels. The DFT band structure, shown as dashed lines, exhibits the interesting features first pointed out by Schoop \textit{et al.} \cite{Schoop-16} and experimentally demonstrated almost simultaneously \cite{Neupane-16}. First is the lifting of degeneracy of the expected Dirac points, along the $\Gamma$-$X/(\nu,0,0)$, $\Gamma$-$M/(\nu,\nu,0)$ and $Z$-$A/\left(\nu,\nu,\frac 12\right)$ symmetry lines. The node along $\Gamma$-$X$ lies on the Fermi energy, and SOC slightly gaps it by 39 meV; gaps with similar amplitudes appear at the other nodes, namely 43 meV along $\Gamma$-$M$ and 41 along $Z$-$A$. As a consequence of the gap opening, the electron (and hole) effective masses are different from, but close to, zero at these notable points, as shown in Table 1. The symmetric topology of the topmost valence band compared to the bottommost conduction band causes electrons and holes to have similar masses, except in the transversal $[001]$ direction; this effect is due to the inherent anisotropy of the ZrSiS crystal structure. Note that electrons and holes are heavier along the transversal directions, significantly for points 1 and 3. Nearly zero effective masses are common to other topological insulators \cite{Narayanan-15}; besides, significant effective mass anisotropy has been identified around the nodal line of ZrSiS \cite{Pezzini-18}. In any case, comparison with experimental data for effective masses is not straightforward, since these are usually indirectly estimated. Hu \textit{et al.} \cite{Hu-17} estimate $m^*$ as 0.025$m_0$ to $0.068m_0$ for ZrSiS from de Haas-van Alphen oscillations for a range of temperatures and magnetic field. Pezzini \textit{et al.} estimate $m^*$ to be larger than $1.0m_e$, with actual values increasing with the magnetic field strength, along given orbits within the nodal line of ZrSiS \cite{Pezzini-18}. These citations cannot be regarded as strong evidences supporting our results, but just as signs that the effective masses of ZrSiS near the gap may be small, as also very anisotropic. 

\begin{table}[!htb]
	\small
	\caption{Effective masses for electron and holes near the gap of ZrSiS, in units of the electron mass $m_0$}
	\begin{tabular*}{0.48\textwidth}{@{\extracolsep{\fill}}lllllll}
		\hline
		    & \multicolumn{3}{l}{Electrons} & \multicolumn{3}{l}{Holes} \\ 
		    & $m_{\parallel}^*$ & $m_{\perp1}^*$ & $m_z^*$ & $m_{\parallel}^*$ & $m_{\perp1}^*$ & $m_z^*$\\ \hline
		    \multicolumn{1}{l}{$\;$Gap at 1$\;$} & $\;0.024\;$ & $\;0.523\;$ & $\;1.944\;$ & $\;0.024\;$ & $\;0.283\;$ & $\;1.944\;$ \\
		    \multicolumn{1}{l}{$\;$Gap at 2$\;$} & $\;0.014\;$ & $\;0.057\;$ & $\;0.850\;$ & $\;0.015\;$ & $\;0.052\;$ & $\;0.680\;$ \\
		    \multicolumn{1}{l}{$\;$Gap at 3$\;$} & $\;0.018\;$ & $\;0.066\;$ & $\;1.047\;$ & $\;0.018\;$ & $\;0.074\;$  & $\;2.268\;$ \\
		 \hline
	\end{tabular*}
\end{table}

The second characteristic is the existence of Dirac-like crossings points at around $-$0.7 eV (with respect to the Fermi energy) at $X\left(\frac 12,0,0\right)$ and $-$0.5 eV at $R\left(\frac 12, 0, \frac 12\right)$. These points are protected by the screw axes of the $P4/nmm$ space group, and are not affected by SOC \cite{Young-15}. Finally, it is noticeable that the valence and conduction bands are linear below and above the gap, respectively. Along $X$ and the gap at point 1 or along $ZR$, the linear regime of the valence band extends down to the Dirac crossing points, but it reaches lowest energies (below $-1.0$~eV) along $\Gamma$ and point 1, or at both sides of points 2 and 3. The bottommost conduction band exhibits a similar topology. 

The $G_0W_0$ corrections to the DFT bands are displayed as solid lines in Figure~\ref{fig:gw_bands}. These corrections preserve the overall band topology; in particular, the Dirac-like nodes are conserved within the accuracy of our procedure ($\pm$~10~meV), while the gaps along $\Gamma$-$X$, $M$-$\Gamma$ and $A$-$Z$ are slightly modified to 60~meV, 70~meV and 60~meV, respectively. Also the Dirac points below the Fermi energy are preserved, slightly shifted downwards though. The most remarkable changes of the band structure appear at $\Gamma$, where the valence band maximum shifts about 1.5~eV below the Fermi level, as well as around the $A$ symmetry point of the BZ. 

\subsection{Optical properties}
\begin{figure}[!htb]
	\centering
	\begin{subfigure}[b]{0.45\textwidth}
		\centering	
		\label{subfig:abs_Ex} 
		\includegraphics[width=0.75\columnwidth]{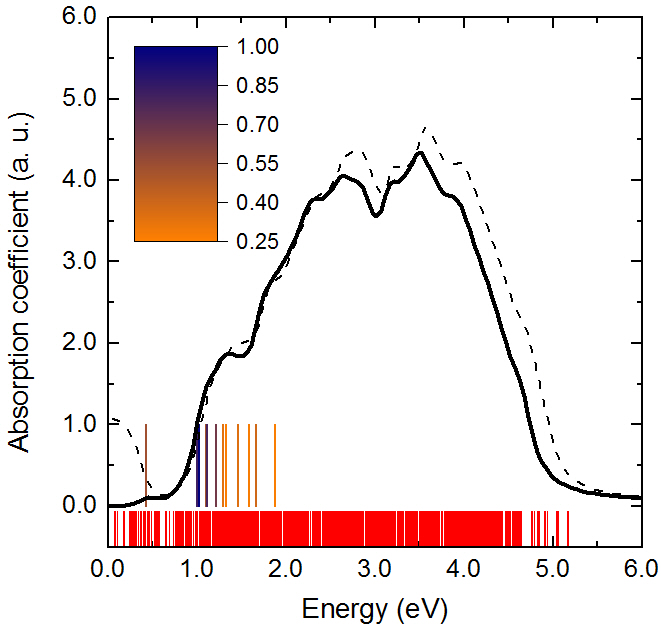}
		\caption{}
	\end{subfigure}

	\vspace{\baselineskip}

	\begin{subfigure}[b]{0.45\textwidth}
		\centering	
		\includegraphics[width=0.75\columnwidth]{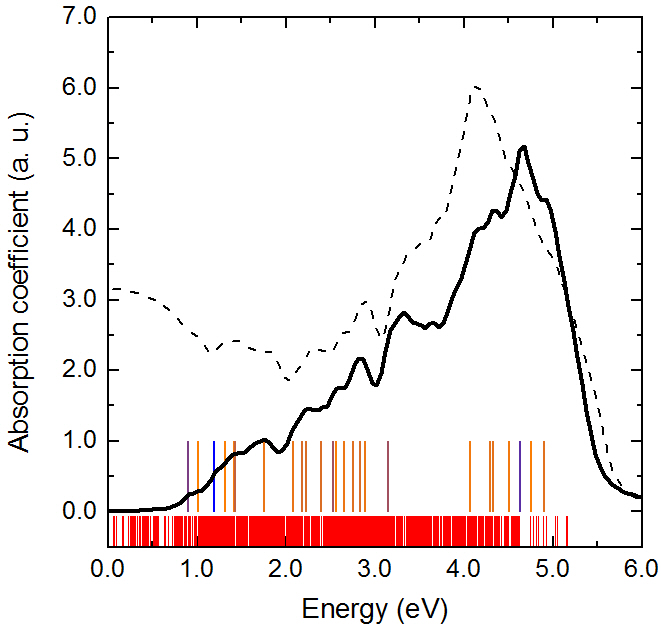}
		\caption{}
	\label{subfig:abs_Ez}
	\end{subfigure}
	\caption{Absorption coefficient vs. photon energy for the electric field parallel to the $Ox$ (\textit a) and the $Oz$ (\textit b) axes. The vertical lines denote the positions of the excitons in the system; see the text for details. Dashed and solid lines correspond to calculations within the IP and BSE approximations, respectively.}
	\label{fig:details}
\end{figure}

Figures~\ref{fig:details} plot the calculated absorption coefficient of $\text{ZrSiS}$ vs. the photon energy for the incident field polarized along the $Ox$ and $Oz$ axes. In these figures, the dashed and solid lines correspond, respectively, to the results obtained within the IP approximation and by solving the BSE. Comparison between IP and BSE results indicate that a significant excitonic activity takes place, as well as that the optical response depends on the polarization of the incident field. The maximum absorption is achieved between 2.0~eV and 4.0~eV for $\bm E \parallel Ox$, that is, within the solar range, and at around 5.0~eV for $\bm E \parallel Oz$. Note that BSE yields a red shift of the absorption spectrum with respect to the IP results for $Ox$ polarization, but a blue shift for $\bm E \parallel Oz$. 

\begin{figure}[!htb]
	\centering
	\includegraphics[width=0.75\columnwidth]{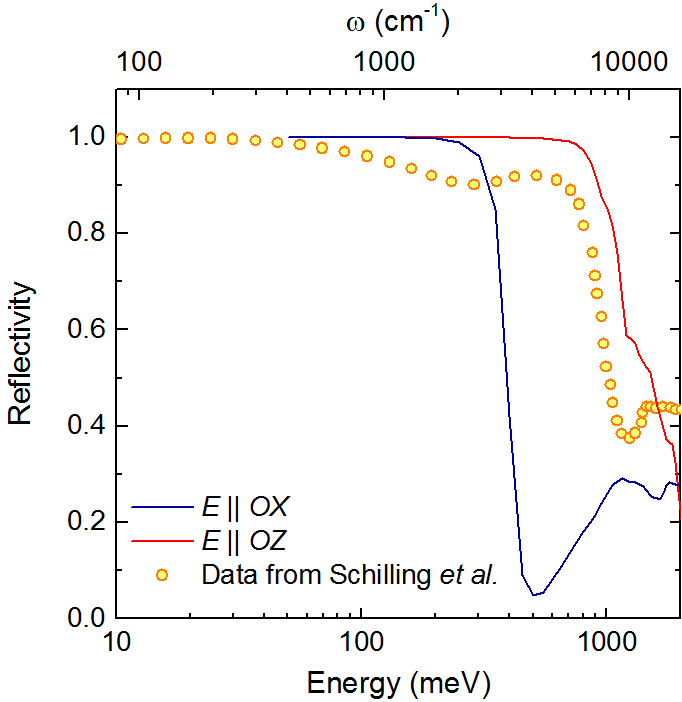}
	\caption{Calculated reflectivity of ZrSiS for different orientations of the incident electric fields. Data from Schilling \textit{et al.} \cite{Schilling-17} are included for comparison.}
	\label{fig:refl}
\end{figure}

The electron-hole interactions result in the formation of a myriad of excitons for both orientations of the field, shown as vertical lines in Figs.~\ref{fig:details}\textit a and \ref{fig:details}\textit b. Most of these are dark, that is, with virtually zero strengths; in this work, excitons with strengths below 10$^{-4}$ are considered dark, and appear as short red vertical lines in Figs.~\ref{fig:details}. For both orientations, the first dark exciton forms along $\Gamma$-$M$ at point 2 of Fig.~\ref{fig:gw_bands} between the topmost valence band and the bottommost conduction band (besides the corresponding degenerate states, of course), and it is unbound. Dark excitons are responsible for the marked difference between the IP and BSE spectra at low energies, namely below 1.0 eV and 2.0 eV for the $Ox$ and $Oz$ orientations, respectively. Indeed, the IP formalism predicts a finite absorption at $\omega \rightarrow 0$ \cite{Zhou-20}, which is in contradiction of the experimental evidence of high reflectivity (virtually total) within that frequency range, which is followed by an abrupt decrease at around 1~eV \cite{Schilling-17}. On the contrary, the BSE formalism yield a low absorption (and therefore a high reflectivity) below 1~eV. Figure~\ref{fig:refl} plots the reflectivity of ZrSiS calculated for both orientations of the electric field, and shows that our theoretical values compare fairly well with the experimental ones [acquired for incidence on (001) surfaces] below 1~eV.

The positions of the bright excitons (those influencing the absorption spectrum) are shown as vertical long lines in Figs.~\ref{fig:details}\textit a and \ref{fig:details}\textit b as well; the color code indicates the relative intensity of each exciton. The number and spectral range of the excitons depend on the polarization of the incident light. For the $Ox$ orientation, bright excitons have energies below 2.0~eV, whereas they extend up to about 5.0~eV for the $Oz$ orientation; note also that the main effect of the electron-hole interactions appear for the latter. The first bright excitons appear virtually at the same energy for both field orientations (namely 0.43~eV and 0.41~eV, for $\bm E$ parallel to $Ox$ and $Oz$, respectively), but their binding energies, respectively 380~meV and 90~meV, depend on the field direction. These excitons appear at different $k-$points between the topmost valence band and the second lowest conduction band. For the $Ox$ polarization it is $\bm k_1 = \left(0,\frac 13, 0.43\right)$, whereas for the $Oz$ polarization the main contribution arises from the state with $\bm k_2 = (0.14,0.29,0)$, with $\bm k_1$ contributing by around 0.5 the intensity at $\bm k_2$. The dissimilarity of the binding energies for the first bright exciton could be explained then by the different effective masses of electrons and holes at the respective $k$ points. The effect of anisotropy on the effective mass of charge carrier has been modeled elsewhere \cite{Schindlmayr-97}. The application of this model to our data indicates that the binding energies for the first excitons for the $E_x$ and $E_z$ polarization should differ by a factor $\approx 8$, which compares reasonably well with the calculated results, given the simplicity of the model. Details of the calculation are given in the Appendix. The most intense excitons, on the other hand, appear at 1.03~eV at 1.20~eV, respectively, for $Ox$ and $Oz$ polarizations, with binding energies 110~meV and 510~meV.  

\begin{figure}[!htb]
	\centering
	\begin{subfigure}[b]{\columnwidth}
		\centering
		\includegraphics[width=0.85\columnwidth]{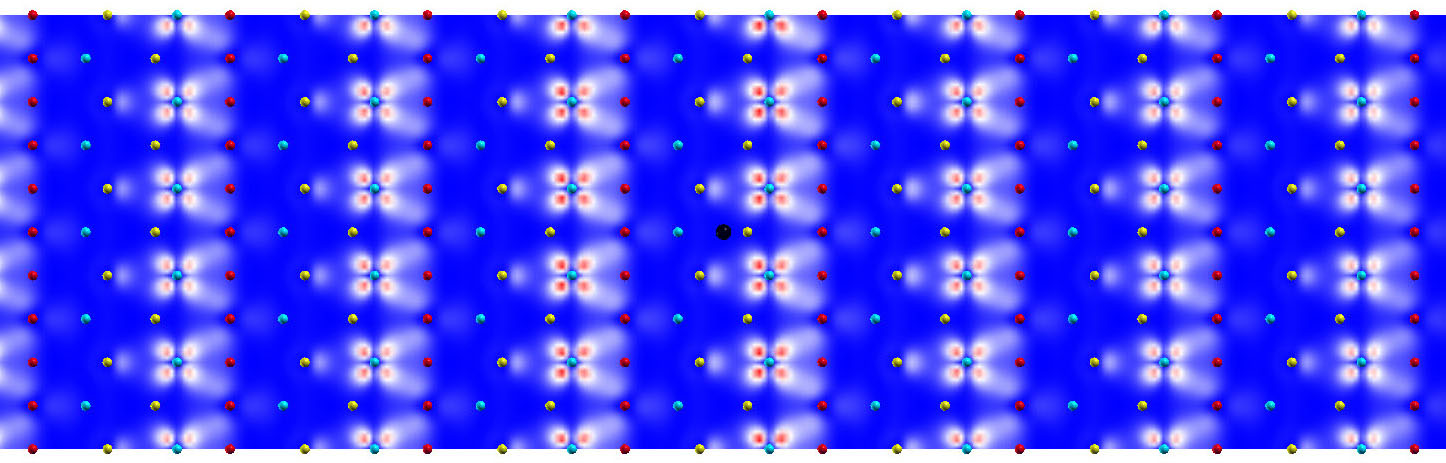}
		\caption{}
	\end{subfigure}
	
	\vspace{\baselineskip}

	\begin{subfigure}[b]{\columnwidth}
		\centering
		\includegraphics[width=0.85\columnwidth]{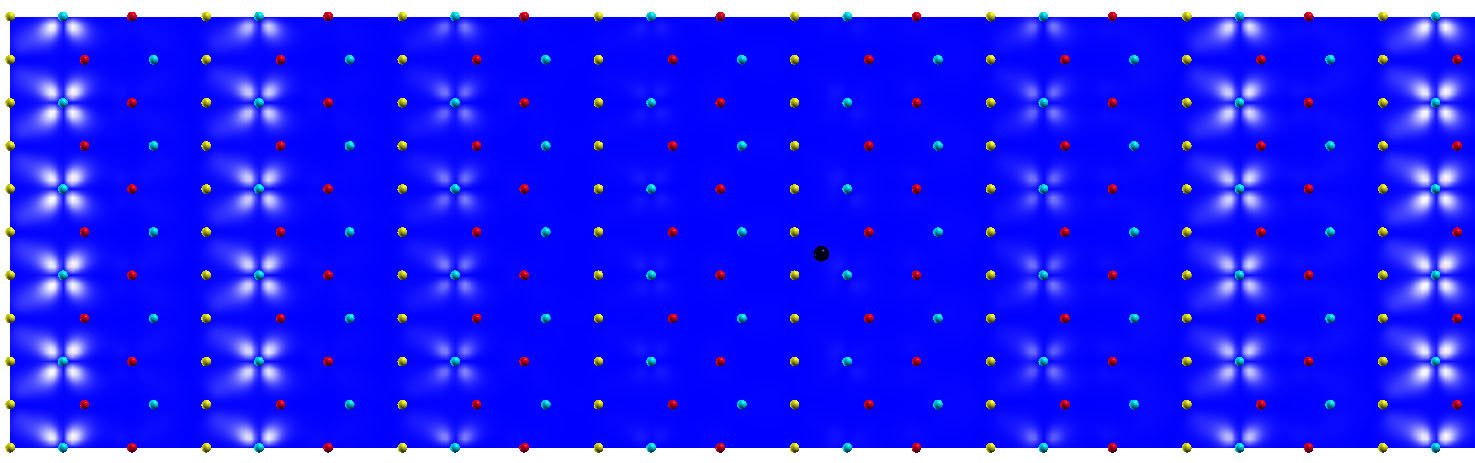}
		\caption{}
	\end{subfigure}
	
	\vspace{\baselineskip}

	\begin{subfigure}[b]{\columnwidth}
		\centering
		\includegraphics[width=0.85\columnwidth]{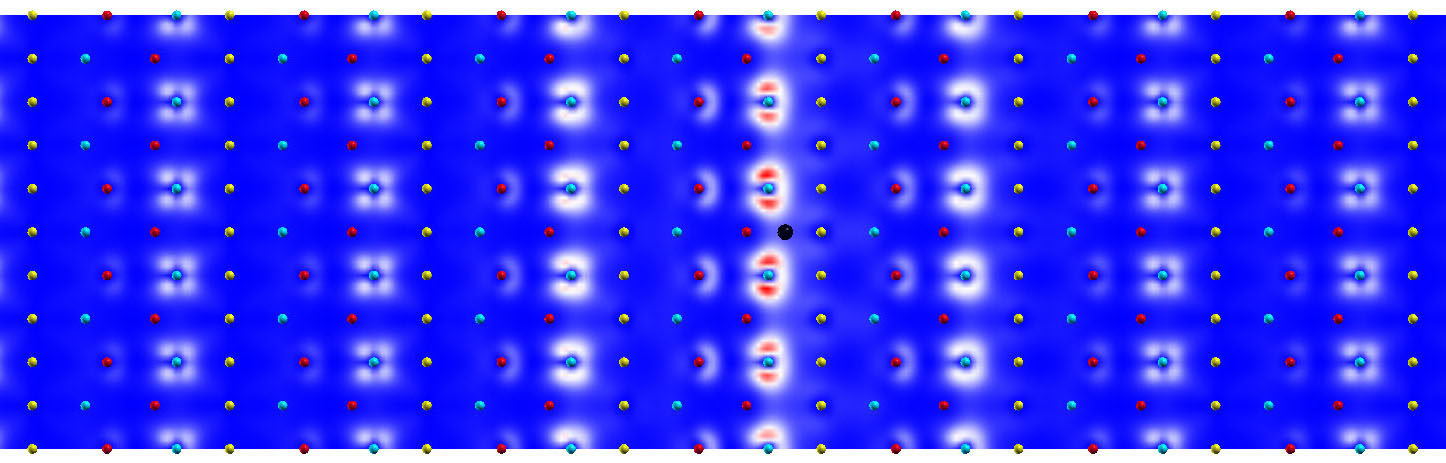}
		\caption{}
	\end{subfigure}
	
	\vspace{\baselineskip}

	\begin{subfigure}[b]{\columnwidth}
		\centering
		\includegraphics[width=0.25\columnwidth]{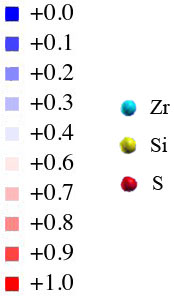}
	\end{subfigure}
	\caption{Spatial distributions of the first dark (\textit a), first bright (\textit b) and most intense (\textit c) excitons for $\bm E \parallel Ox$.}		
	\label{fig:charge_Ex}
\end{figure}

\begin{figure}[!htb]
	\centering
	\begin{subfigure}[b]{\columnwidth}
		\centering
		\includegraphics[width=0.85\columnwidth]{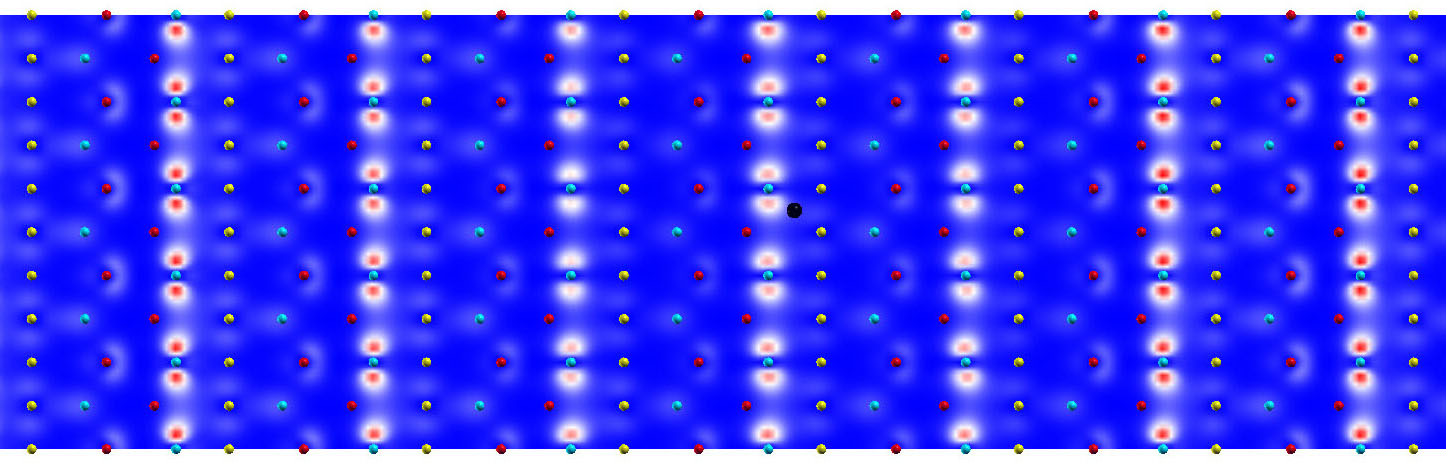}
		\caption{}
	\end{subfigure}
	
	\vspace{\baselineskip}

	\begin{subfigure}[b]{\columnwidth}
		\centering
		\includegraphics[width=0.85\columnwidth]{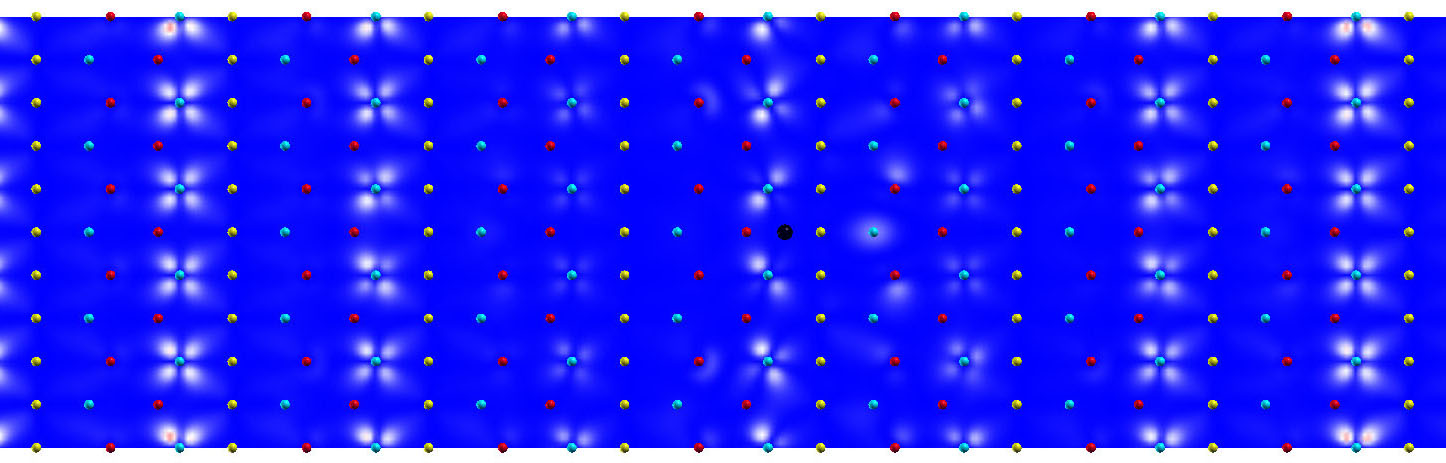}
		\caption{}
	\end{subfigure}
	\caption{First bright (\textit a) and most intense (\textit b) excitons for $\bm E \parallel Oz$. The color code is that of Figure~\ref{fig:charge_Ex}}
	\label{fig:charge_Ez}
\end{figure}

Figures~\ref{fig:charge_Ex} plot the spatial distribution (i.e., the probability of electron location for a given position $\bm r_{h,0}$
of the hole) for the first dark (\textit a), first bright (\textit b) and most intense (\textit c) excitons for the electric field parallel to $Ox$. Analogously, Figures~\ref{fig:charge_Ez} plot the first bright (\textit a) and most intense (\textit b) excitons for $\bm E \parallel Oz$. These distributions are calculated from the excitonic states as
\begin{equation}
	\rho^{\lambda}(\bm r_e) = \sum_{c,v,\bm k} |\langle c(\bm r_e)\,v(\bm r_{h,0})\,\bm k|\lambda\rangle|^2,
\end{equation}
where the sum spans the DFT states contributing to the state $|\lambda\rangle$. In all cases, the hole was placed at a maximum of the charge density. The formation of the first dark exciton for $Ox$ polarization involves just a slight distortion of the electronic density near the hole, which is visible in Figure~\ref{fig:charge_Ex}\textit a as blue spots in the $\pi$ orbitals at the Si - S bonds and Zr atoms at cells adjacent to the hole. This is consistent with the fact that unbound excitons have wide spacial extensions. Bright excitons give rise to more important charge redistributions, in particular for $\bm E \parallel Ox$ by virtue of its larger binding energy. In this case, the excitons extend about three unit cells, with the electronic charge redistributing roughly similar to the dark excitons. For $\bm E \parallel Oz$, the spatial extension of the exciton is quite modest. In opposition to the previous case, the electronic charge now arranges mostly at $\sigma$ orbitals. 

\subsection{Plasmonic activity}\label{sec:plasmon}
\begin{figure}[!htb]
	\centering
	\begin{subfigure}[b]{\columnwidth}
		\centering
		\includegraphics[width=0.85\columnwidth]{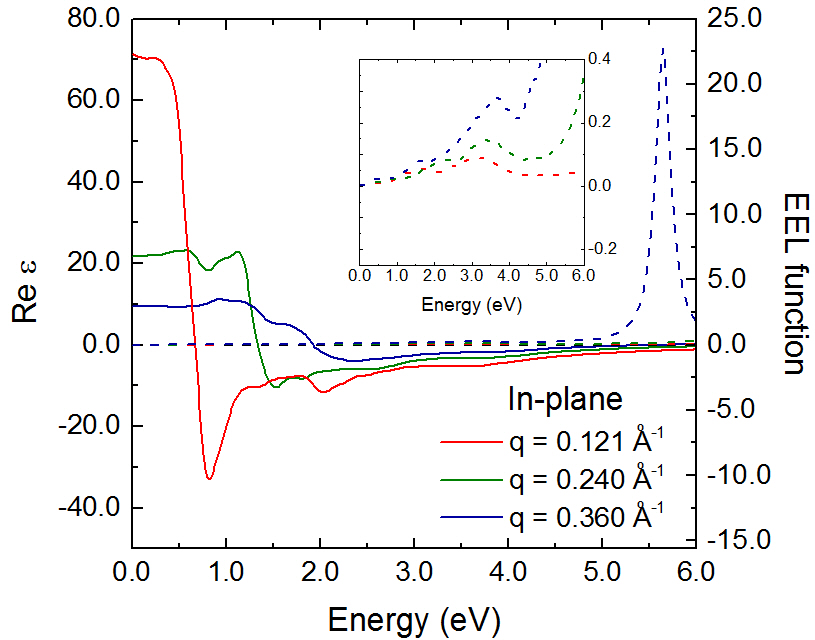}
		\caption{}		
	\end{subfigure}
	
	\vspace{\baselineskip}

	\begin{subfigure}[b]{\columnwidth}
		\centering
		\includegraphics[width=0.85\columnwidth]{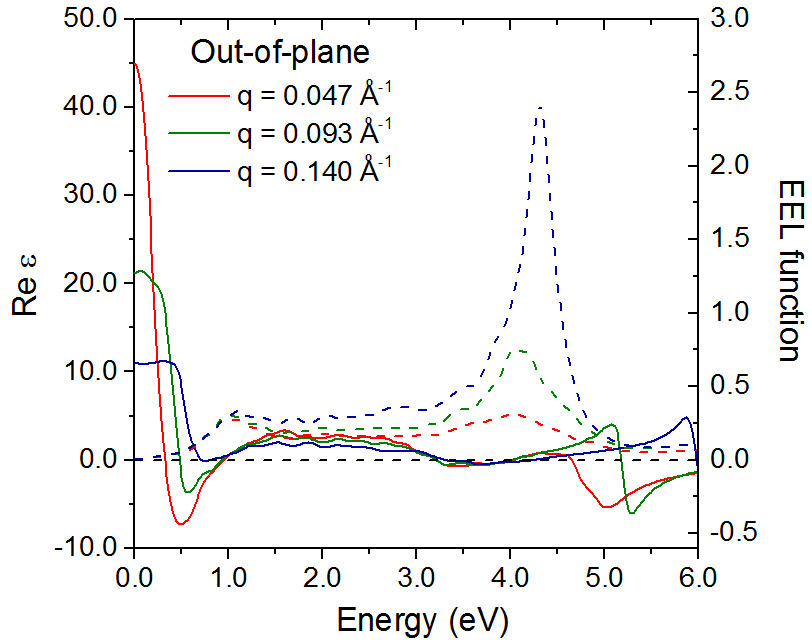}
		\caption{}
		\end{subfigure}
	\caption{Re $\varepsilon(\bm q,\epsilon)$ (left axes, solid lines) and $L(\bm q, \epsilon)$ (right axes, dashed lines) as functions of energy for in-plane (\textit a) and out-of-plane (\textit b) scattering $\bm q$ vectors. The inset in (\textit a) shows a detail of the $L(\bm q, \epsilon)$ functions. Only data for $\bm E \parallel Ox$ are shown; data for $\bm E \parallel Oz$ are virtually identical.}
	\label{fig:epsilon}
\end{figure}

An interesting question which remains to be clarified is the possible existence of plasma excitations, which was studied by Zhou et al. from an independent-particle approach \cite{Zhou-20}. Plasma excitations may propagate at frequency ranges for which the real part of the dielectric functions becomes negative; therefore, the frequency ranges of interest are defined by the condition $\text{Re }\varepsilon(\bm q, \omega)=0$. Zhou et al. identified two regions of possible plasma oscillations, namely 5-7~eV, which are damped and therefore hardly detectable, and 19-20~eV, which could be experimentally observed. Figures~\ref{fig:epsilon} plot the Re $\varepsilon(\bm q, \omega)$ function computed within the BSE formalism for several $\bm q$ in-plane and out-of-plane vectors (solid lines) for $\bm E \parallel Ox$ within the low-energy regime of Zhou et al. Our solutions of the BSE indicate that the frequency ranges for possible plasmons depend on the $\bm q$ vector. For the in-plane geometry, the upper limit for the plasmon frequency is around 6 eV, the lower one increasing from 0.7~eV to 1.9~eV as $q$ increases. For the out-of-plane geometry, the real part of the permittivity becomes negative at two regimes, namely between 0.3-0.9~eV and 4.7-6~eV and 0.5-0.8~eV and 5.2-6~eV for $q=0.047$\text{ \AA}$^{-1}$ and $q=0.093$\text{ \AA}$^{-1}$, respectively, and nowhere for $q=0.140$\text{ \AA}$^{-1}$ and beyond. 

\begin{figure}[!htb]
	\centering
	\begin{subfigure}[b]{\columnwidth}
		\centering
		\includegraphics[width=0.75\columnwidth]{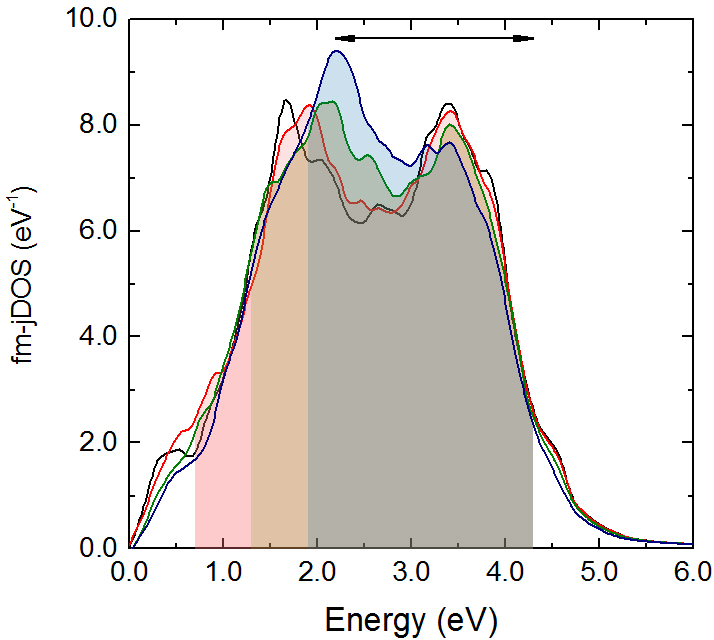}
		\caption{}
	\end{subfigure}
	
	\vspace{\baselineskip}

	\begin{subfigure}[b]{\columnwidth}
		\centering
		\includegraphics[width=0.75\columnwidth]{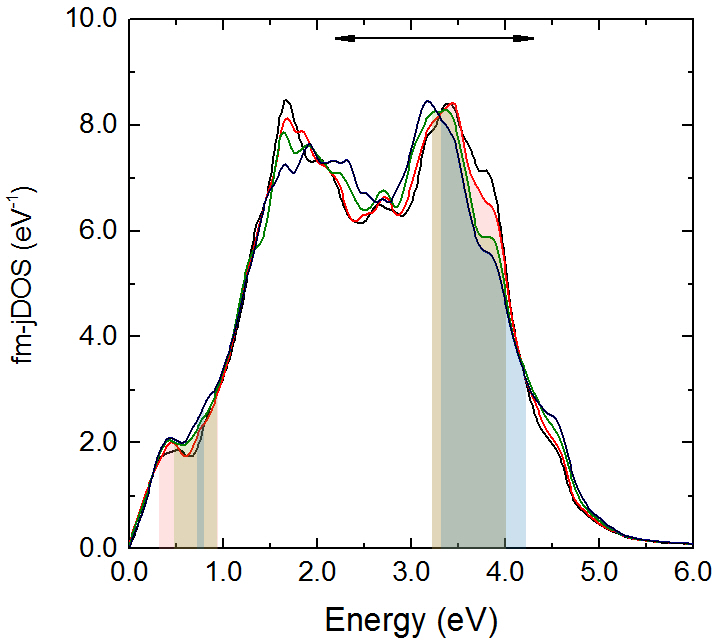}
		\caption{}
	\end{subfigure}
	\caption{Finite-momentum jDOS vs. energy for $\bm E \parallel Ox$ for in-plane (\textit a) and out-of-plane (\textit b) conditions, with the same color code as in Figure~\ref{fig:epsilon}. The shaded regions correspond to the plasmon energies. The double arrow denote the decay range for $\bm q = \bm 0$ (optical limit).}
	\label{fig:jdos}
\end{figure}

The nature and characteristics of the plasmons may be investigated from the electron energy loss and finite-moment joint density of states functions, defined in \eqref{eq:def_L} and \eqref{eq:def_j} above. The right axes of Figures~\ref{fig:epsilon} plot the $L(\bm q, \omega)$ functions for the in-plane (\textit a) and out-of-plane (\textit b) geometries (dashed lines). These plots evidence that $L(\bm q, \omega)$ takes on negligible values within the energy ranges where plasmons propagate, which indicates that these are actually damped. Thus, the BSE formalism confirms the previous IP predictions that no plasmons propagate at energies below 6~eV in bulk ZrSiS. In fact, these plasmons decay into electron-hole pairs. Figures~\ref{fig:jdos} show the fm-jDOS for $\bm E \parallel Ox$ for $\bm q \perp Oz$ (\textit a) and $\bm q \parallel Oz$; the shaded regions indicate the energy range of plasmons for each condition. These figures show that the generation of electron-hole pairs is important for $\bm q \perp Oz$. These pairs differ by vectors of the form $(\nu, \nu, 0)$, and the decay rate becomes higher, but within a narrower energy range, as $q$ increases. Identifying these pairs on the $G_0W_0$ band structure of Fig.~\ref{fig:gw_bands} is virtually impossible due to the high number of possible transitions involved, but we notice that one of the possible pairs $(\bm q, \omega)$ satisfying the conservation laws is that linking point 1 at the valence band to the Dirac point at $X$ at the conduction band. By contrast, pair generation is modest for $\bm q \parallel Oz$ due to that the energy range for plasmons is narrower. 

\begin{figure}[H]
	\centering
	\includegraphics[width=0.75\columnwidth]{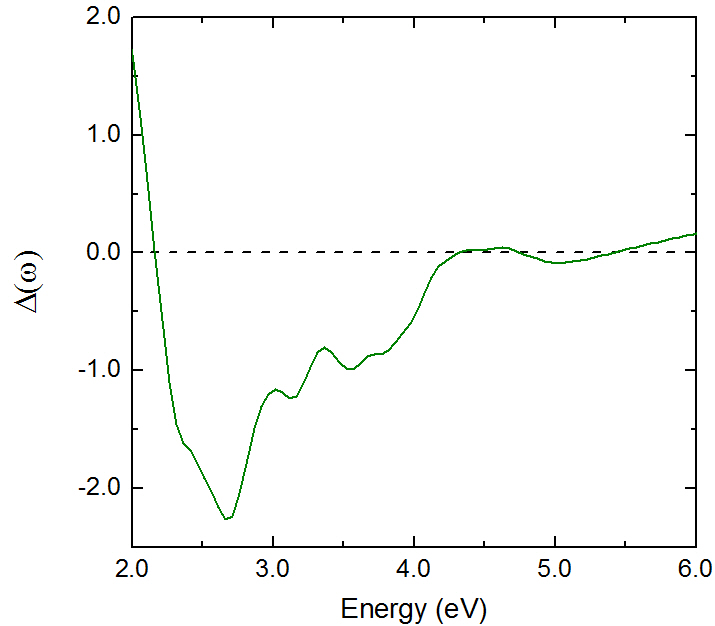}
	\caption{Re [$\varepsilon_{xx} (\omega)] \cdot$ Re $[\varepsilon_{zz} (\omega)]$ vs. energy for ZrSiS within the BSE approximation.}
	\label{fig:s1}
\end{figure}

Two final comments are relevant. First is that BSE calculations yield no significant differences for the two orientations of the incident electric field regarding plasmon activity; this also agrees with the IP prediction that plasmonic behavior is independent of the polarization of light. Besides, it is remarkable that BSE confirms the existence of hyperbolic plasmons, which were previously theorized from the IP approach \cite{Shekhar-14, Gomez-15}. Hyperbolic plasmons appear in highly anisotropic solids in which the sign of the effective permittivity changes with respect to the incident electric field. In a tetragonal system, propagation of light is ruled by the relation
\begin{equation}
	\frac {k_x^2+k_y^2}{\varepsilon_{zz}(\omega)} + \frac {k_z^2}{\varepsilon_{xx}(\omega)} = \frac {\omega^2}{c^2},
	\label{eq:sup_6}
\end{equation}

\noindent where $\varepsilon_{xx}(\omega)$ and $\varepsilon_{zz}(\omega)$ are the in-plane and out-ot-plane permittivities, respectively. Eq.~(\ref{eq:sup_6}) yields a hyperboloid for 

\begin{equation}
	\varepsilon_{xx}(\omega) \cdot \varepsilon_{zz}(\omega) < 0 
	\label{eq:hyp_plasmons}
\end{equation}
which defines the condition for which hyperbolic plasmons may propagate. Figure~\ref{fig:s1} plots $\Delta(\omega) \equiv$ Re [$\varepsilon_{xx} (\omega)] \cdot$ Re $[\varepsilon_{zz} (\omega)]$ as a function of energy within the range 2.0$-$6.0 eV. According to this figure, $\Delta <$~0 within a range 2.1 eV wide centered at 3.3~eV, approximately, and also within 0.6 eV around 5.0 eV, approximately. The latter is consistent the results obtained within the IP approximation \cite{Zhou-20}, but the former arises as a consequence of the electron-hole interaction. Note, however, that modes within the respective energy ranges are damped, which indicates that hyperbolic plasmons are unlikely to be detectable. 
\smallskip

\section*{Acknowledgments}
Financial support by Grants FIS2016-76359-P (Ministry of Economy and Competitiveness of Spain), RTI2018-093711-B-I00 (Ministry of Science of Spain), IB16013 (Junta de Extremadura, Spain) and Prometeo GVA2020-16 (Generalitat Valenciana, Spain) are greatly acknowledged. Thanks are also given to the Spanish Supercomputing Network for the use of the supercomputer Tirant and to the Center for Nanoscale Materials (Advance Photon Source, Argonne National Lab.) for the approval of proposal \#69398.

\bibliography{zrsis}
\bibliographystyle{rsc}

\section*{Appendix}
Let us consider a solid with a gap whose constant energy surfaces $\varepsilon(\bm k)$ are ellipsoids, with effective masses $m_\parallel$ and $m_\perp$ for electrons and holes along the ellipsoid axis and in the plane perpendicular to it, respectively. In such a system, the binding energy of an exciton is given by
\begin{equation}
	E(\lambda) = -\frac 32 \left(\frac {e^2}{4\pi\varepsilon\hbar}\right)^2 \left(\frac 2{\mu_\perp} + \frac 1{\lambda^2\mu_\parallel}\right)^{-1}I(\lambda)^2, \tag{A1}
	\label{eq:sup_1}
\end{equation}
where $e$ is the electron charge, $\varepsilon$ is the dielectric constant of the solid, $\hbar$ is the reduced Planck's constant and $\mu_\parallel$ and $\mu_\perp$ are the reduced masses parallel and perpendicular to the ellipsoid axis, respectively. The function $I(\lambda)$ is given by

\begin{widetext}
\begin{equation}
	I(\lambda) = \frac 12\int_0^{\pi} \frac {\sin \theta d\theta}{\sqrt{1+(\lambda^2-1)\cos^2\theta}} = \begin{dcases} \dfrac {\asinh \sqrt{\lambda^2-1}}{\sqrt{\lambda^2-1}} \text{   for }\lambda > 1 \\ \dfrac {\arcsin \sqrt{1-\lambda^2}}{\sqrt{1-\lambda^2}} \text{   for }\lambda < 1  \end{dcases} \tag{A2}
\end{equation}
\end{widetext}

and $\lambda$ is the solution of

\begin{equation}
	\frac {\mu_\perp}{\mu_\parallel} = 2\lambda^2 \frac {1-\lambda I(\lambda)}{I(\lambda)-\lambda} \tag{A3}
\end{equation}

Full details about this formalism can be found in \cite{Schindlmayr-97}. In our case, the effective masses for the first excitons are summarized in table below. According to this table, the binding energies for $\bm E \parallel Ox$ and $\bm E \parallel Oz$ are $0.02E_0$ and $0.16E_0$, with $E_b =  -\dfrac 32 \left(\dfrac {e^2}{4\pi\varepsilon\hbar}\right)^2$. These values indeed differ by a factor 8, as the main text indicates.

\begin{widetext}
\begin{center}
\begin{table}[!htb]
	\label{tabla}
	\centering
	\def\arraystretch{1.1}
	\begin{tabular}{|c | c c c c | c | c | c |}
		\hline \hline
		\multicolumn{8}{|c|}{$\bm E \parallel Ox$} \\ \hline \hline
		 & $m_{\perp,1}$ & $m_{\perp,2}$ & $m_{\perp} = \sqrt{m_{\perp,1}m_{\perp,2}}$ & $m_\parallel$ & $\mu_\perp$ & $\mu_\parallel$ & $\lambda$ \\ \hline
		 Electrons & 0.363 & 0.218 & 0.281 & 0.011 & \multirow{2}{*}{0.223} & \multirow{2}{*}{0.010} & \multirow{2}{*}{3.1417}\\
		 Holes & 4.611 & 0.257 & 1.089 & 0.136 & & & \\ \hline \hline
		 
		\multicolumn{8}{|c|}{$\bm E \parallel Oz$} \\ \hline \hline
		 & $m_{\perp,1}$ & $m_{\perp,2}$ & $m_{\perp} = \sqrt{m_{\perp,1}m_{\perp,2}}$ & $m_\parallel$ & $\mu_\perp$ & $\mu_\parallel$ & $\lambda$ \\ \hline
		 Electrons & 0.609 & 3.718 & 1.505 & 1.134 & \multirow{2}{*}{0.576} & \multirow{2}{*}{0.378} & \multirow{2}{*}{1.162} \\
		 Holes & 0.261 & 3.342 & 0.933 & 0.567 & & & \\ \hline \hline
	\end{tabular} 
\end{table}
\end{center}
\end{widetext}

\end{document}